# Influence of Ti$^{3+}$ Defect-type on Heterogeneous Photocatalytic H$_2$ Evolution Activity of TiO$_2$

*Shiva Mohajernia[a], Pavlina Andryskova[b], Giorgio Zoppellaro[c], Seyedsina Hejazi[a], Stepan Kment[c], Radek Zboril[c], and Patrik Schmuki[a,c,d]\**

[a] Department of Materials Science and Engineering WW4-LKO University of Erlangen-Nuremberg Martensstrasse-7, Erlangen D-91058, Germany

[b] Department of Biophysics, Faculty of Science, Palacký University Olomouc, 17. Listopadu 12, 771 46 Olomouc, Czech Republic

[c] Regional Centre of Advanced Technologies and Materials, Palacky University Olomouc, 17. listopadu 50A, 772 07 Olomouc, Czech Republic

[d] Chemistry Department, Faculty of Sciences, King Abdulaziz University, 80203, Jeddah, Saudi Arabia

\* Corresponding author. E-mail: schmuki@ww.uni-erlangen.de

**Abstract:**

Reduced titanium dioxide has recently attracted large attention, particularly for its unique co-catalyst-free $H_2$ heterogeneous photocatalytic application. The enhanced photocatalytic activity of the reduced $TiO_2$ was previously ascribed to the introduction of point crystal defects (mainly $Ti^{3+}$ centers), which result in the formation of intrinsic co-catalytic centers and enhanced visible light absorption. In this work, we systematically investigate the effect of different defects in the $TiO_{2-x}$ lattice on photocatalytic $H_2$ evolution. To introduce different types of defects, thermal annealing in air (oxidative), Ar (inert), Ar/$H_2$ (reducing), and $H_2$ (reducing) atmospheres were performed on commercially available anatase nanopowder. Then, the powders were characterized by scanning electron microscopy (SEM), X-ray diffraction (XRD) and high-resolution transmission electron microscopy (HR-TEM) to clarify the effect of treatment on material properties. Furthermore, the defect types were characterized by electron paramagnetic resonance (EPR) spectroscopy. We show that thermal annealing in different atmospheres can form different amounts of different defect types in the $TiO_2$ structure. The highest photocatalytic activation is achieved by annealing the anatase powder in a reducing atmosphere for an appropriate temperature/annealing time. By combining the results from $H_2$ generation and EPR analysis we show that the simultaneous presence of two types of defects, i.e. surface exposed $Ti^{3+}$ and lattice embedded $Ti^{3+}$ centers, in an optimum low concentration, is the determining factor for an optimized photocatalytic $H_2$ evolution rate. In fact, annealing anatase powder under the so-reported optimized conditions in reducing atmosphere leads to the generation of a considerable amount of $H_2$, with rates as high as 338 µmolh$^{-1}$g$^{-1}$.

**Introduction:**

Titanium dioxide ($TiO_2$), and particularly its anatase polymorph, is still the most investigated semiconductor for photocatalytic $H_2$ generation by virtue of its exceptional photocorrosion resistance, low cost, and suitable band-edge positions for $H_2$ formation from $H_2O$ [1–3]. Thermodynamically, photo-generated conduction band electrons, ejected by UV irradiation, are able to reduce $H_2O$ to $H_2$. However, due to a sluggish kinetics of electron transfer from $TiO_2$ to electrolyte, and owing to trapping and recombination of charge carriers, the efficiency of such a photocatalytic system is very low. [2,4,5]

To tackle these limitations in charge transfer, co-catalysts, mostly noble metal nanoparticles (NPs) such as Pt, Au and Pd, are used [5,6]. They act as electron sinks and facilitate efficient electron-transfer across the interface. However, due to the high cost of noble metals, alternative approaches could be based on intrinsic doping by charge transfer mediators, i.e. the formation of suitable $Ti^{3+}$ centers, introduced by oxygen vacancies (VOs) in $TiO_2$ lattice. Reports on the successful creation of such species are mainly centered on exposing $TiO_2$ structures to a high-temperature treatment in various reducing atmospheres (e.g., vacuum, Ar, Ar/$H_2$, and pure $H_2$).[7–10]

In 2011 Chen et al. reported for the first time that upon a $H_2$ thermal treatment at 200 °C under high pressure, a defective structure is formed (black $TiO_2$) and has a high photocatalytic activity for $H_2$ generation if Pt is used as a co-catalyst. The authors attributed the enhanced performance of black $TiO_2$ to the increased light absorption due to the narrowing of the optical band gap of black $TiO_2$ to 1.54 eV. [1] Later on, in 2014 Liu et al. reported on significant open circuit $H_2$ evolution from grey $TiO_2$ nanotubes and from grey anatase, without the use of any noble metal as co-catalyst.[8,11] Liu et al. ascribed the high open-circuit hydrogen generation of the hydrogenated anatase to the formation of $Ti^{3+}$ species which act as cocatalytic centers facilitating electron transfer to the solution.[11] Different methods of $TiO_2$ reduction, such as high energy ion implantation, [12] electrochemical reduction, [13] and hydride ball milling, [14] have then been reported to lead to hydrogen generation without using any noble metals. Liu et al. found that photocatalytic hydrogen evolution of the black $TiO_2$ is mainly ascribed to the specific defect formation and not coupled with their visible light absorption behavior.[8–10,15,16] This finding was further investigated in more detail and confirmed by Mohajernia et al. studying the effect of visible light absorption versus conductivity on the enhanced photoelectrochemical performance of reduced $TiO_2$ nanotubes.[15] Measuring the open-circuit hydrogen evolution of reduced $TiO_2$ nanotubes under visible irradiation, the authors proved the established visible light absorption hardly contributes at all to $H_2$ generation.[15] The aforementioned studies clearly establish that $Ti^{3+}$ species play a crucial role in photocatalysis of titanium sub-oxides. The reported active material have in common that (the key) $Ti^{3+}$ is present only in very low concentration. It is, however, well known that detection of $Ti^{3+}$ species in minute amount is challenging with common characterization techniques such as XRD, XPS, TEM or EDX. An outstanding technique to elucidate on different types of $Ti^{3+}$ centers is the use of an electron spin sensitive technique. The most precise method for investigating the alterations in terms of detecting the presence and type of paramagnetic species in reduced $TiO_2$ structure is electron paramagnetic resonance spectroscopy (EPR). Reduction of $TiO_2$ associated with removal of oxygen atoms from the lattice results in formation of paramagnetic $Ti^{3+}$ sites. Based on previous EPR studies on defective $TiO_2$, the main spin containing centers usually formed in $TiO_2$ from synthesis are a) $Ti^{3+}$ centers located in regular lattice positions, b) surface exposed $Ti^{3+}$, c) interstitial $Ti^{3+}$, and d) oxygen ions with a trapped hole (oxygen-based radicals)[8,11,16–19]. Although several studies observed the formation of various signals of $Ti^{3+}$ in reduced samples, there is no systematic study that correlates the types of spin-containing defects with the photocatalytic activity of reduced $TiO_2$ structures [8,11,16,20–22].

In the present work, we performed such a systematic investigation on a variety of reduced $TiO_2$ materials and critically discuss the possible correlation between number (spin concentration) and types of point defects with the experimentally observed photocatalytic activity of reduced $TiO_2$ anatase powders. Point defects were formed intentionally under inert (Ar), reductive (Ar/$H_2$ and $H_2$) and oxidative (air) conditions and at different temperatures, in order to vary both the defect identity and the total amount. We observe that a subtle balance of embedded and surface exposed defects in low concentration does have a large impact on enhancing the photocatalytic activity of $TiO_2$ nanostructures. This balance of defects can be adjusted by annealing anatase powder under mild temperature conditions and in reducing atmosphere; the balanced presence of surface exposed and lattice embedded $Ti^{3+}$ sites in low concentration leads to a considerable amount of $H_2$, with rate as high as 338 µmolh$^{-1}$g$^{-1}$ under 365 nm illumination. On the contrary, the introduction of a high concentration of spin containing sites obtained from the use of reducing atmospheres under harsh conditions is detrimental for the photocatalytic performance. Similar detrimental results were also obtained when oxidative or inert annealing conditions were employed.

**Result and discussion**

Figure 1a shows optical images of anatase powders thermally treated in air, Ar, Ar/$H_2$ and $H_2$ environments at various temperatures for 1 hour. The images illustrate that the color of the powders treated in air and Ar atmosphere remain almost unchanged, while an obvious color transition from white to black for the powders reduced in Ar/$H_2$ and $H_2$ atmosphere from 300°C to 900°C is evident. Figure S1a-d compares the corresponding absorbance spectra of the thermally treated anatase samples. Figure 1b compares the spectrum in different atmosphere at a fixed temperature of 500°C. The absorption spectra are well in line with color changes in optical images and. as anticipated, show high absorbance in the UV range, corresponding to the bandgap of $TiO_2$ (3.2 eV for anatase and ~ 3.0 eV for rutile). The absorbance spectra of the powders treated at Ar-500, Ar/$H_2$-500 and $H_2$-500 show a tail in the visible range, which is in general most significant for the $H_2$ and Ar/$H_2$ treated samples. Although the absorption results may suggest visible photo-activity of the darker nanopowders, this point needs to probed by photocatalytic measurement in visible range, which will be discussed in the photocatalysis section of this paper. In Figure S1e-h the absorption spectra are re-plotted according to the Kubelka-Munk equation. Band gap evaluation results show that the $TiO_2$ powder annealed in oxidative (air) and inert (Ar) environment show a band-gap decrease that we can ascribe to the transformation of anatase to rutile phase at elevated temperature (700°C) as verified by the XRD (below). Noteworthy, for the powders treated in reducing atmosphere (Ar/$H_2$ and pure $H_2$) this transformation is postponed, and at 900°C no band gap can be evaluated for these samples. This can be ascribed to phase changes to semimetalic phases [23], as later confirmed by XRD.

Figure 1 c-f and Figure S2-5 show the SEM images of the anatase powder treated at different temperatures and environments. Figure 1 c-f indicate that by thermal treatment of the anatase nanopowders up to 500°C in Ar/$H_2$ atmosphere, the powder size remains intact, while by further increasing the annealing temperature, the nanopowders start sintering which results in a significantly increased $TiO_2$ powder size (even particles with diameter >500 nm can be observed after annealing at 900°C – see Fig. S6). Also in other atmospheres a similar trend was observed. The SEM images in Figure S3-5 show a comparison of annealing environment. The nature of the environment does not have a significant effect on particle size, whereas temperature plays a crucial role in sintering – see particle average size in Figure S7. This is in line with previous reports on pressure-less sintering behavior of $TiO_2$ powders [24,25]. TEM images in Figure S8 also confirm the particle size dependency on temperature and show the sintering effect at elevated temperatures. It has been reported that reduced $TiO_2$ structures show in some cases an amorphous shell around a crystalline core. To investigate the formation of this amorphous shell, high-resolution TEM was made for a core sample in this report that is the 500°C Ar/$H_2$ treated powder; however, no amorphous shell was detected (Figure S9).

XRD measurements were carried out on pristine and thermally treated $TiO_2$ nanopowders in Ar and Ar/$H_2$ environment and are presented in Figure 1g,h. The diffractograms identify anatase phase of $TiO_2$ (JCPDS Card no. 21-1272) in the pristine sample and the samples which were annealed below 500°C; while by increasing the annealing temperature conversion to rutile phase (JCPDS Card no.71-0650) is initiated. However, dependent on the annealing environment, the temperature and extent of transition from anatase to rutile phase varies, i.e. the conversion from anatase to rutile is facilitated (occurs at a lower temperature) in oxidative (air) and inert environment (Ar) as compared to a reductive environment (Ar/$H_2$ and $H_2$) (see Figure 1c,d and Figure S10 a,b). To compare the phase content in different environments and different temperatures more precisely, Rietveld analysis was performed and the data are summarized in Table S1. The quantitative data confirm the facilitated transformation of anatase to rutile in oxidative and inert environment. Thus, SEM, XRD, and TEM results jointly confirm that powder morphology and crystallographic structures (anatase) all remain unchanged by annealing at temperatures below 500°C.

In order to study the photocatalytic behavior of the different thermally treated $TiO_2$ powders, open circuit photocatalytic $H_2$ evolution was measured using 2 mg powder dispersed in water/ methanol (50:50) solution and illuminated by 365 nm LED. Figure 2a shows the comparison of $H_2$ evolution rate of the pristine anatase powder and the powders treated in air, Ar, $H_2$ and $Ar/H_2$ atmosphere in different temperatures. The results show that air as well as Ar annealing treatment do not have any significant impact on the open circuit $H_2$ evolution and it is almost identical to the amount of $H_2$ generated from the non-treated anatase powder. Therefore, the small amount of $H_2$ evolved from these samples is ascribed to the self-activation of the anatase powder under UV illumination – this effect was previously investigated thoroughly by Wierzbicka et al. [26]. However, for the thermally treated powders in reducing environment at 500 °C, the $H_2$ generation rates are significantly enhanced. For higher annealing temperature, however, they steadily decay. This decay at elevated temperature can be observed for all the samples, regardless of annealing environment. In order to determine the contribution of visible light absorption (c.f. Figure 1b) in photocatalytic $H_2$ evolution of pristine and thermally treated powders, photocatalytic $H_2$ evolution was measured under 450 nm laser irradiation (see Figure 2b). The results do not show any photocatalytic activity in the visible range. In the UV range, it is worth mentioning that samples reduced in $Ar/H_2$ and $H_2$ show a very similar trend and slightly higher hydrogen evolution rate for the sample which was treated at 500°C in $H_2$ environment, in contrast to earlier results [27].

To study the effect of treatment time, anatase powder was treated at 500°C for 15 min to 960 min in $Ar/H_2$ (see Figure 2c). The results show that photocatalytic $H_2$ evolution rate increased for the samples treated in $Ar/H_2$ up to 480 min; however, by further increasing the time of thermal treatment in reducing atmosphere, the photocatalytic activity started dropping significantly. Please note that for all samples XRD shows no change in crystallinity.

To understand the difference in nature of the defect types and how they can affect the photocatalytic activity of the treated powders, electron paramagnetic resonance spectroscopy (EPR) was employed at X-band frequency (9.15-9.16 GHz). Figure 3a-d show the EPR results of the air, Ar, Ar/$H_2$ and $H_2$ treated powders recorded at 123 K. Table S2 collects the estimated g-tensor values from EPR analysis associated to the various spin-containing sites, and includes evaluation of the total spin concentration as obtained from comparison with $CuSO_4$ spin standard. From Figure 3a-d, all $TiO_2$ samples annealed at 300°C under various atmospheres exhibit similar EPR resonances, with low intensity signals arising from lattice embedded $Ti^{3+}$ sites. These signals show the usual *g* tensor values known for these $Ti^{3+}$ spin containing species, falling in the range 1.99-1.97 for the *g*-perpendicular (x,y) component [28]. The recorded EPR signatures suggest that the low temperature treatment does not lead to the formation of a significant defective structure in the $TiO_2$ system. By increasing the annealing temperature to 500°C, significant differences start to emerge in response to the diverse atmosphere used during annealing. When $TiO_2$ is annealed under air (Figure 3a, Air-500 trace), besides the presence of lattice embedded $Ti^{3+}$ resonance, a broad EPR resonance arising from $Ti^{3+}$ sites that are surface exposed appears clearly (around g = 1.93), together with signals arising from trapped holes (at g > 2.0023) [17].

In Ar atmosphere and at 500°C (Figure 3b), multiple signatures from lattice embedded $Ti^{3+}$ sites exhibiting different rhombic distortions evolved, with $g_{x,y}$ components falling at 1.998 and 1.993. Moreover, an unusual and strong hole signature appears in the low field region, featuring $g_z$ = 2.011, $g_y$ = 1.971, $g_x$ = 1.921. These g-tensor values are rather unusual; we speculate that such signature arises from an oxygen-based radical (hole) being trapped well below the nanoparticle's surface and spatially located close to the $Ti^{3+}$ spin containing site. These spin containing sites do not contribute in enhancing the proclivity of the system in the photocatalytic $H_2$ production (see Figure 2a) but rather appears to hamper the electron transfer process to $H^+$. Similar hole species exhibiting comparable g-tensor values ($g_z$ = 2.010, $g_y$ = 1.970, $g_x$ = 1.921) emerge also in the $TiO_2$ sample treated in air, but only when the higher annealing temperature is employed (Fig. 3a, annealing temperature 700 °C, Air-700 trace) and remain present as well in the Ar sample annealed at 700 °C (Figure 3c, Ar-700 trace). In the Supporting information (Figure S11), as an example an EPR simulated resonance envelope is given (calculated in the spin-Hamiltonian framework by perturbation theory approximation) of the air-700 sample, from which it becomes evident that an unusually large contribution stems from hole species (34% contribution of the total spin density). The $Ar/H_2$ and $H_2$ samples annealed at 500 °C (Figure 3c and Figure 3d) show very different EPR resonance signatures compared to the Ar treated counterpart (Figure 3b) and appear rather similar to the Air-500 sample (Figure 3a). In $Ar/H_2$ and $H_2$, the resonance features observed are attributed to $Ti^{3+}$ sites that are lattice embedded ($g_{x,y}$ around 1.99), to hole centers (e.g. surface $Ti^{IV}$-$O^{\bullet}$ centers, with g values > 2.0023) and $Ti^{3+}$ sites that are surface exposed (with $g_{avg}$ around 1.93). The major difference between $Ar/H_2$ and $H_2$ samples annealed at 500 °C is the larger amount of spin containing defects that are generated in $H_2$ at 500 °C ($5.33 \times 10^{18}$ spin/g, see also Table S2 and Figure 3e), compared to those estimated in the $Ar/H_2$ ($3.00 \times 10^{18}$ spin/g). Increasing the temperature up to 700 °C shows that the samples annealed in air and in Ar behave again similarly to each other, exhibiting both the dominant presence of the hole signature. However, while the total spin density increases upon increasing the temperature in the Air sample (from $2.00 \times 10^{18}$ spin/g at 500°C to $4.52 \times 10^{18}$ spin/g at 700 °C), for the Ar sample the reverse process occurs (from $3.52 \times 10^{18}$ spin/g at 500°C to $1.52 \times 10^{18}$ spin/g at 700 °C). A further increase in the temperature up to 900 °C induces a total loss in the EPR resonance signals in the Air sample and leaves a very small amount of spin containing sites in the Ar sample ($0.47 \times 10^{18}$ spin/g). As for $Ar/H_2$ and $H_2$ treated $TiO_2$ powder, their behavior at high annealing temperature (700-900 °C) is very different to both Air and Ar treated samples, showing that when a more reducing environment is used and higher annealing temperatures are reached, a boost in the generation of spin containing sites is observed. In fact, the Ar/H$_2$ and H$_2$ treated samples at 700 °C and 900 °C (Figure 3c and Figure 3d) show very similar EPR resonance signatures, namely they display a strong, broad, and unresolved resonance line around g$_{avg}$ = 1.95, with total spin density that significantly increases upon temperature increase (at 700 °C and 900 °C, see Table S2 and Figure 3e). Taking together these results from EPR analysis, some additional information on the temperature induced generation of spin containing defects in these TiO$_2$ materials can be drawn. The temperature dependency of vacancy generated in a crystalline material can be expressed by Eq.(1):

$$N_V = N \exp(-Q_V / \kappa_B T) \qquad \text{Eq. (1)}$$

where $N_V$ is the vacancy concentration, $Q_V$ is the energy required for vacancy formation, $k_B$ is the Boltzmann constant, $T$ is the absolute temperature, and $N$ is the concentration of atomic sites. This equation is valid for a crystalline and solid material and is regardless of annealing environment, i.e., by increasing the temperature the number of vacancies increases exponentially to stabilize the increase in the entropy of the crystal. This is clearly occurring in all samples tested, as validated from spin density estimation, in the temperature range 300 °C to 500 °C (see Table S2 and Figure 3e). The process of additional defects generation should continue until it starts to interfere with a physical phase transition (melting, or a solid-state phase transition). Taking the case of air and Ar annealed samples, embedded oxygen vacancies increase until a rutile transformation induces a drop in the number of defects (see Figure 3a and 3b) with almost complete loss at 900 °C.

For the samples treated in the Ar/H$_2$ and H$_2$ environment the EPR results and spin concentration estimation show that with increasing temperature, in addition to the enhancement of embedded defects which can be explained by Eq. 1, the signals from the surface exposed vacancies are also increased significantly. This is well in line with the previous proposed mechanism of defect formation in a reducing environment which state that reducing agent (in this case $H_2$) will react with the surface exposed oxygen atoms and eliminate the oxygen atoms from the lattice surface [29].

Intense EPR signals of the surface exposed $Ti^{3+}$ for the powder treated at 900°C in $Ar/H_2$ and $H_2$ reveals that by increasing the annealing temperature, this type of spin containing sites become the dominant species ($g_{avg}$ ~1.95). Based on the EPR results, it seems that the high number of defects present in reducing atmosphere at high temperature indeed prevents formation of photogenerated charge separation and favors, on the contrary, their fast recombination (probably a coulomb effect, due to the presence of an already large electric field all over the material).

In order to remove the impact of rutile formation that arises from an increase in the annealing temperature, we reduced the samples at 500°C but using different time intervals. Figure S10c show the XRD result of the powder treated in $Ar/H_2$ at 500°C in different time intervals. XRD results indicate anatase phase for all the samples treated for different times. Interestingly, EPR results show a very similar trend of transformation from embedded defects to fully surface exposed defects for the sample treated for longer times at 500°C in $Ar/H_2$ (See Figure S12). Combining the time and temperature dependent results we suggest that in a reducing environment the number of surface exposed defects and embedded defects can be tuned by changing the time and temperature of the annealing. Figure S13 provides a schematic image of the surface exposed and embedded oxygen vacancy ($Ti^{3+}$) sites. For simplicity purposes, the crystals are presented along the 001 and 010 directions.

Comparing the EPR signals of the optimized samples in terms of photocatalytic activity in inert and reducing atmosphere and also the EPR spectra of the sample treated in $Ar/H_2$ in an optimized time, we suggest that a low density mixture of embedded and surface exposed $Ti^{3+}$ sites seems vital for achieving the highest photocatalytic activity, while it is detrimental to pursue high amount of defects induced. However, it is still unclear if such low-density admixture of spin containing defects are directly entering, as active sites, in the hydrogen photoproduction or, rather, their presence brings in a more efficient generation of other photoactive spin containing species during turn-over. Nevertheless, the present work shows clearly the significance of optimized defect engineering toward the "right" combination that is the key pre-requisite to generate photocatalytically active sites for $H_2$ generation.

**Conclusion**

In summary, we provide a systematic investigation of anatase and the introduced defects as a result of thermal treatment in an inert, an oxidative and in a reductive environments using EPR spectroscopy. We demonstrate that reducing $TiO_2$ powders in different environments, temperatures and time intervals leads to powders with a wide range of optical and electronic characteristics. More importantly, the various thermally treated powders contain distinctly different defect types. Based on the EPR results, we prove that the presence of an optimized level of two types of defects, namely surface exposed $Ti^{3+}$ and embedded $Ti^{3+}$ is the key to obtain a drastically enhanced photocatalytic $H_2$ generation from $TiO_2$ surfaces.

**Experimental**

Anatase powder samples were purchased from Sigma Aldrich (purity 99.8%, particle size: 25–35 nm). Thermal treatment of the powders was carried out by annealing in air, Ar and Ar/$H_2$ 10% at different temperatures. For open-circuit photocatalytic $H_2$ evolution measurements, the powder (2 mg) was dispersed in 10 mL oxygen-free aqueous methanol (50 vol.%) solution to prepare a suspension in a quartz tube. The suspension was illuminated either under 365 nm LED. A gas chromatograph (GCMS-QO2010SE, SHIMADZU) with TCD detector was used to measure the amount of $H_2$ generated. The absorbance of the samples was measured between 200–1000 nm (Avantes, AvaLight-DH-S-BAL, using cable IC-DB26-2). The blank quartz glass was used as reference. A Teflon standard was used as background. The band gap evaluation was performed using Kubelka-Munk equation:

$$[F(R)h\nu]^{\frac{1}{2}} = A(h\nu - E_g)$$

where A is a constant, F(R) is the Kubelka-Munk function and Eg is the band gap.

The morphology of the samples was characterized by field-emission scanning electron microscopy (Hitachi S4800). Transmission electron microscopy was carried out with a Philips CM30 TEM. X-ray diffraction (XRD), performed with an X'pert Philips MPD (equipped with a Panalytical X'celerator detector) using graphite monochromized Cu Kα radiation (λ = 1.54056 Å), was used to analyze the crystallographic properties of the materials. Rietveld Analysis was performed on XRD diffractograms using a MAUD software by Luca Lutterotti[30].

EPR spectra of the $TiO_2$ powder samples were recorded on JEOL JES-X-320 operating at X-band frequency (~9.15-.16 GHz), equipped with a variable temperature control ES 13060DVT5 apparatus. The cavity Q quality factor was kept above 6000 in all measurements and signal saturation was avoided by working under low-applied microwave power (applied $P_a$ << 1.0 mW). Highly-pure quartz tubes were employed (Suprasil, Wilmad, ≤ 0.5 OD). The EPR spectra were all recorded by using the following parameters: modulation frequency 100 kHz, modulation amplitude 0.5 mT, time constant 0.03 s, sweep time of 4 min and temperature $T$ = 123 K. Experimental determination of the spin concentration in the various $TiO_2$ preparations was obtained by using $CuSO_4 \times 5\ H_2O$ (99.999%, CAS Number: 7758-99-8) as $S$=1/2 standard and used in powder form, by comparison of the reciprocal double integrated resonance signals divided by square root of the applied power.

**Acknowledgement:**

The authors would like to acknowledge the ERC, the DFG and the Operational program research, Development and education (European regional development Fund, Project No. CZ.02.1.01/0.0/0.0/15_003/0000416 of the ministry of education, youth and sports of the Czech Republic) for financial support.

**References**

[1]   X. Chen, L. Liu, P.Y. Yu, S.S. Mao, Increasing Solar Absorption for Photocatalysis with Black Hydrogenated Titanium Dioxide Nanocrystals, Science (80-. ). 331 (2011) 746–750. http://science.sciencemag.org/content/331/6018/746.abstract.

[2]   A. Fujishima, T.N. Rao, D.A. Tryk, Titanium dioxide photocatalysis, J. Photochem. Photobiol. C Photochem. Rev. 1 (2000) 1–21. doi:https://doi.org/10.1016/S1389-5567(00)00002-2.

[3]   R. Asahi, T. Morikawa, T. Ohwaki, K. Aoki, Y. Taga, Visible-Light Photocatalysis in Nitrogen-Doped Titanium Oxides, Science (80-. ). 293 (2001) 269–271. http://science.sciencemag.org/content/293/5528/269.abstract.

[4]   F.E. Osterloh, Inorganic Materials as Catalysts for Photochemical Splitting of Water, Chem. Mater. 20 (2008) 35–54. doi:10.1021/cm7024203.

[5]   X. Chen, S.S. Mao, Titanium dioxide nanomaterials: Synthesis, properties, modifications and applications, Chem. Rev. 107 (2007) 2891–2959. doi:10.1021/cr0500535.

[6]   J. Prakash, S. Sun, H.C. Swart, R.K. Gupta, Noble metals-TiO2 nanocomposites: From fundamental mechanisms to photocatalysis, surface enhanced Raman scattering and antibacterial applications, Appl. Mater. Today. 11 (2018) 82–135. doi:https://doi.org/10.1016/j.apmt.2018.02.002.

[7]   P. Jonsen, Identification of different hydrogen-reduced titania crystallographic forms by1H NMR spectroscopy, Catal. Letters. 2 (1989) 345–349. doi:10.1007/BF00768176.

[8]   N. Liu, X. Zhou, N.T. Nguyen, K. Peters, F. Zoller, I. Hwang, C. Schneider, M.E. Miehlich, D. Freitag, K. Meyer, D. Fattakhova-Rohlfing, P. Schmuki, Black Magic in Gray Titania: Noble-Metal-Free Photocatalytic H2 Evolution from Hydrogenated Anatase, ChemSusChem. 10 (2017) 62–67. doi:10.1002/cssc.201601264.

[9]   S. Mohajernia, S. Hejazi, A. Mazare, N.T. Nguyen, I. Hwang, S. Kment, G. Zoppellaro, O. Tomanec, R. Zboril, P. Schmuki, Semimetallic core-shell TiO2 nanotubes as a high conductivity scaffold and use in efficient 3D-RuO2 supercapacitors, Mater. Today Energy. 6 (2017) 46–52. doi:https://doi.org/10.1016/j.mtener.2017.08.001.

[10]  N. Liu, C. Schneider, D. Freitag, E.M. Zolnhofer, K. Meyer, P. Schmuki, Noble-Metal-Free Photocatalytic H2 Generation: Active and Inactive 'Black' TiO2 Nanotubes and Synergistic Effects, Chem. - A Eur. J. 22 (2016) 13810–13814. doi:10.1002/chem.201602714.

[11]  N. Liu, C. Schneider, D. Freitag, M. Hartmann, U. Venkatesan, J. Müller, E. Spiecker, P. Schmuki, Black TiO2 Nanotubes: Cocatalyst-Free Open-Circuit Hydrogen Generation, Nano Lett. 14 (2014) 3309–3313. doi:10.1021/nl500710j.

[12]  N. Liu, V. Häublein, X. Zhou, U. Venkatesan, M. Hartmann, M. Mačković, T. Nakajima, E. Spiecker, A. Osvet, L. Frey, P. Schmuki, "Black" TiO 2 Nanotubes Formed by High-Energy Proton Implantation Show Noble-Metal- co -Catalyst Free Photocatalytic H 2 -Evolution, Nano Lett. 15 (2015) 6815–6820. doi:10.1021/acs.nanolett.5b02663.

[13] J.-W. Yun, K.Y. Ryu, T.K. Nguyen, F. Ullah, Y. Chang Park, Y.S. Kim, Tuning optical band gap by electrochemical reduction in TiO2 nanorods for improving photocatalytic activities, RSC Adv. 7 (2017) 6202–6208. doi:10.1039/C6RA25274E.

[14] X. Zhou, N. Liu, J. Schmidt, A. Kahnt, A. Osvet, S. Romeis, E.M. Zolnhofer, V.R.R. Marthala, D.M. Guldi, W. Peukert, M. Hartmann, K. Meyer, P. Schmuki, Noble-Metal-Free Photocatalytic Hydrogen Evolution Activity: The Impact of Ball Milling Anatase Nanopowders with TiH2, Adv. Mater. 29 (2017) 1604747. doi:10.1002/adma.201604747.

[15] S. Mohajernia, S. Hejazi, A. Mazare, N.T. Nguyen, P. Schmuki, Photoelectrochemical H2 Generation from Suboxide TiO2 Nanotubes: Visible-Light Absorption versus Conductivity, Chem. – A Eur. J. 23 (2017) 12406–12411. doi:10.1002/chem.201702245.

[16] A. Naldoni, M. Altomare, G. Zoppellaro, N. Liu, Š. Kment, R. Zbořil, P. Schmuki, Photocatalysis with Reduced TiO2: From Black TiO2 to Cocatalyst-Free Hydrogen Production, ACS Catal. 9 (2019) 345–364. doi:10.1021/acscatal.8b04068.

[17] M. Chiesa, M.C. Paganini, S. Livraghi, E. Giamello, Charge trapping in TiO2 polymorphs as seen by Electron Paramagnetic Resonance spectroscopy, Phys. Chem. Chem. Phys. 15 (2013) 9435–9447. doi:10.1039/C3CP50658D.

[18] C.P. Kumar, N.O. Gopal, T.C. Wang, M.-S. Wong, S.C. Ke, EPR Investigation of TiO2 Nanoparticles with Temperature-Dependent Properties, J. Phys. Chem. B. 110 (2006) 5223–5229. doi:10.1021/jp057053t.

[19] J.M. Coronado, A.J. Maira, J.C. Conesa, K.L. Yeung, V. Augugliaro, J. Soria, EPR Study of the Surface Characteristics of Nanostructured TiO2 under UV Irradiation, Langmuir. 17 (2001) 5368–5374. doi:10.1021/la010153f.

[20] M. Wajid Shah, Y. Zhu, X. Fan, J. Zhao, Y. Li, S. Asim, C. Wang, Facile Synthesis of Defective TiO2−x Nanocrystals with High Surface Area and Tailoring Bandgap for Visible-light Photocatalysis, Sci. Rep. 5 (2015) 15804.

[21] Y. Duan, M. Zhang, L. Wang, F. Wang, L. Yang, X. Li, C. Wang, Plasmonic Ag-TiO2−x nanocomposites for the photocatalytic removal of NO under visible light with high selectivity: The role of oxygen vacancies, Appl. Catal. B Environ. 204 (2017) 67–77. doi:https://doi.org/10.1016/j.apcatb.2016.11.023.

[22] M. Fittipaldi, D. Gatteschi, P. Fornasiero, The power of EPR techniques in revealing active sites in heterogeneous photocatalysis: The case of anion doped TiO2, Catal. Today. 206 (2013) 2–11. doi:https://doi.org/10.1016/j.cattod.2012.04.024.

[23] S. Mohajernia, S. Hejazi, A. Mazare, N.T. Nguyen, I. Hwang, S. Kment, G. Zoppellaro, O. Tomanec, R. Zboril, P. Schmuki, Semimetallic core-shell $TiO_2$ nanotubes as a high conductivity scaffold and use in efficient 3D-$RuO_2$ supercapacitors, Mater. Today Energy. 6 (2017). doi:10.1016/j.mtener.2017.08.001.

[24] D. Li, S. Chen, Y. Jing, W. Shao, Y. Zhang, W. Luan, The master sintering curve for pressureless sintering of TiO2, Sci. Sinter. (2007). doi:10.2298/SOS0702103L.

[25] V.N. Koparde, P.T. Cummings, Sintering of titanium dioxide nanoparticles: a comparison between molecular dynamics and phenomenological modeling, J. Nanoparticle Res. 10

(2008) 1169–1182. doi:10.1007/s11051-007-9342-3.

[26] E. Wierzbicka, X. Zhou, N. Denisov, J. Yoo, D. Fehn, N. Liu, K. Meyer, P. Schmuki, Self-Enhancing H2 Evolution from TiO2 Nanostructures under Illumination, ChemSusChem. 12 (2019) 1900–1905. doi:10.1002/cssc.201900192.

[27] N. Liu, C. Schneider, D. Freitag, U. Venkatesan, V.R.R. Marthala, M. Hartmann, B. Winter, E. Spiecker, A. Osvet, E.M. Zolnhofer, K. Meyer, T. Nakajima, X. Zhou, P. Schmuki, Hydrogenated Anatase : Strong Photocatalytic Dihydrogen Evolution without the Use of a Co-Catalyst ** Angewandte, (2014) 14425–14429. doi:10.1002/ange.201408493.

[28] D.C. Hurum, A.G. Agrios, K.A. Gray, T. Rajh, M.C. Thurnauer, Explaining the Enhanced Photocatalytic Activity of Degussa P25 Mixed-Phase TiO2 Using EPR, J. Phys. Chem. B. 107 (2003) 4545–4549. doi:10.1021/jp0273934.

[29] X. Pan, M.-Q. Yang, X. Fu, N. Zhang, Y.-J. Xu, Defective TiO2 with oxygen vacancies: synthesis, properties and photocatalytic applications, Nanoscale. 5 (2013) 3601–3614. doi:10.1039/C3NR00476G.

[30] L. Lutterotti, Total pattern fitting for the combined size–strain–stress–texture determination in thin film diffraction, Nucl. Instruments Methods Phys. Res. Sect. B Beam Interact. with Mater. Atoms. 268 (2010) 334–340. doi:https://doi.org/10.1016/j.nimb.2009.09.053.

**Figure Captions:**

**Figure 1.** (a) Optical images of different powders thermally treated in air, Ar, $H_2$ and $Ar/H_2$ at different temperatures; (b) Absorbance spectra of the pristine anatase and powders treated at 500º C in air, Ar, $H_2$ and $Ar/H_2$; SEM images of the (c) anatase, (d) 500º C $Ar/H_2$ treated powder, (e) 700º C $Ar/H_2$ treated powder, (f) 900º C $Ar/H_2$ treated powder, (g) XRD patterns of the Ar treated powders at different temperature; (h) XRD diffractograms of the $Ar/H_2$ treated powders at different temperatures.

**Figure 2.** (a) Comparison of the open circuit $H_2$ evolution rate of the anatase powders treated in different atmospheres at various temperatures, (b) comparison of the $H_2$ evolution rate of powders treated at optimum temperature of 500º C in different atmospheres under different light source (UV and visible) irradiation, (c) $H_2$ evolution rate of the optimum powder (500º C $Ar/H_2$) thermally treated at different time intervals.

**Figure 3.** X-band (9.15-9.16 GHz) EPR spectra recorded at $T$ = 123 K of the $TiO_2$ nanoparticles treated under different atmosphere compositions and under different annealing temperatures. (a) air treated $TiO_2$ powder subjected to 300,500,700,900 ºC annealing temperatures, (b) Ar treated $TiO_2$ powder subjected to 300,500,700,900 ºC annealing temperatures, (c) $Ar/H_2$ treated $TiO_2$ powder subjected to 300,500,700,900 ºC annealing temperatures, (d) $H_2$ treated $TiO_2$ powder subjected to 300,500,700,900 ºC annealing temperatures. (e) The experimentally determined spin concentration (spin/g) *vs* annealing temperature ($T$, ºC) of the various $TiO_2$ materials shown in panels (a)-(d).

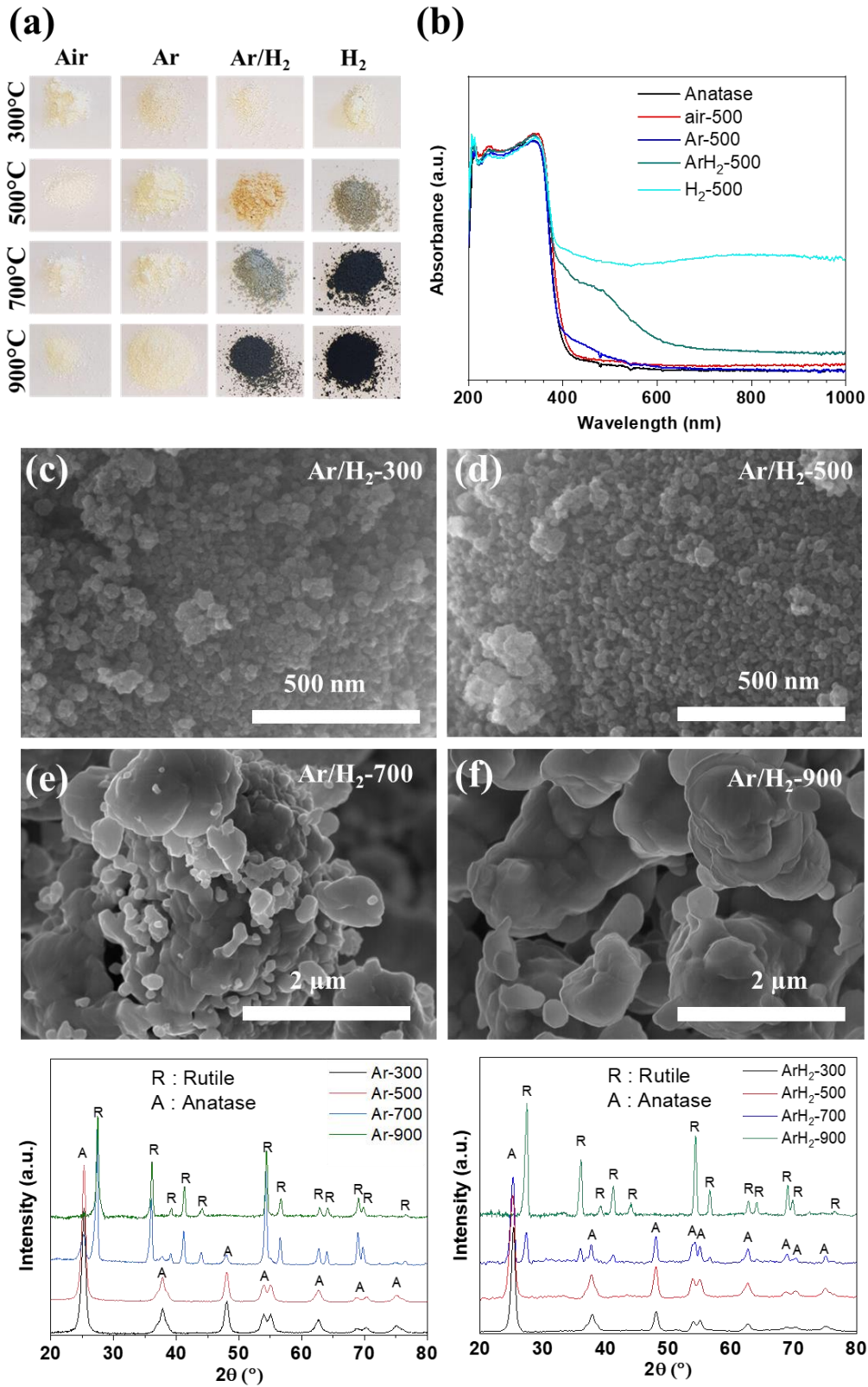

**Fig.1**

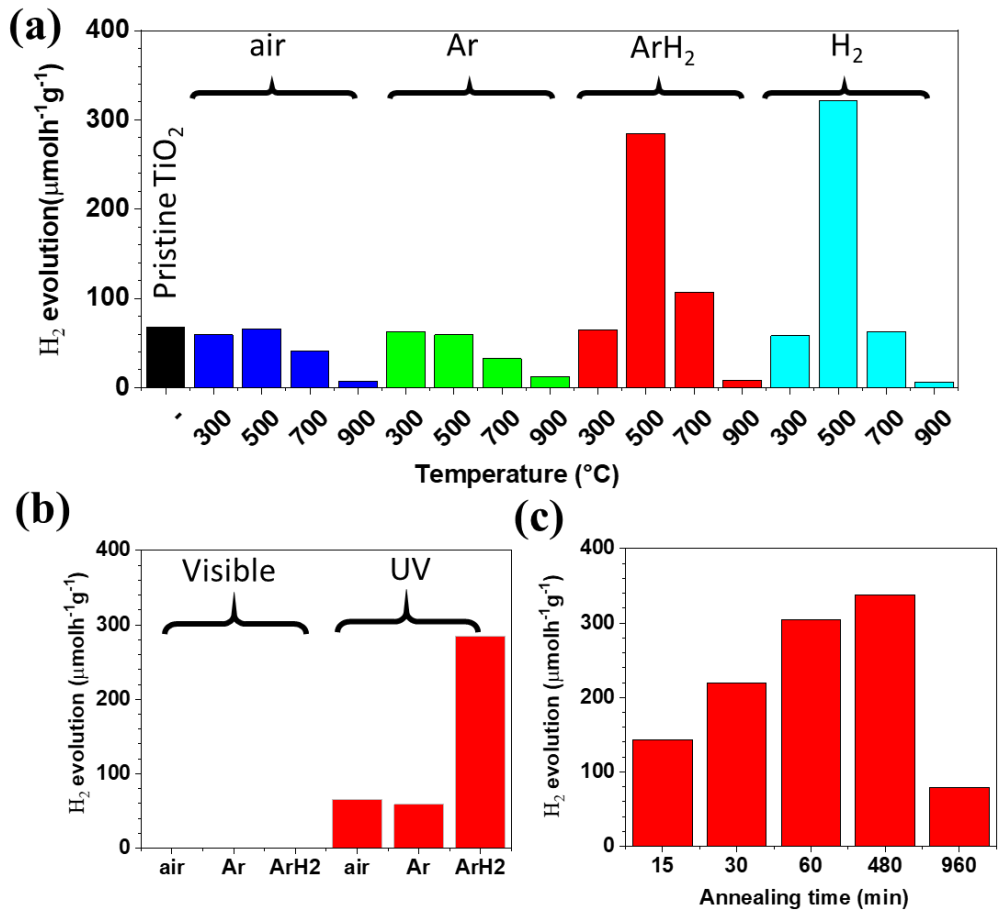

**Fig.2**

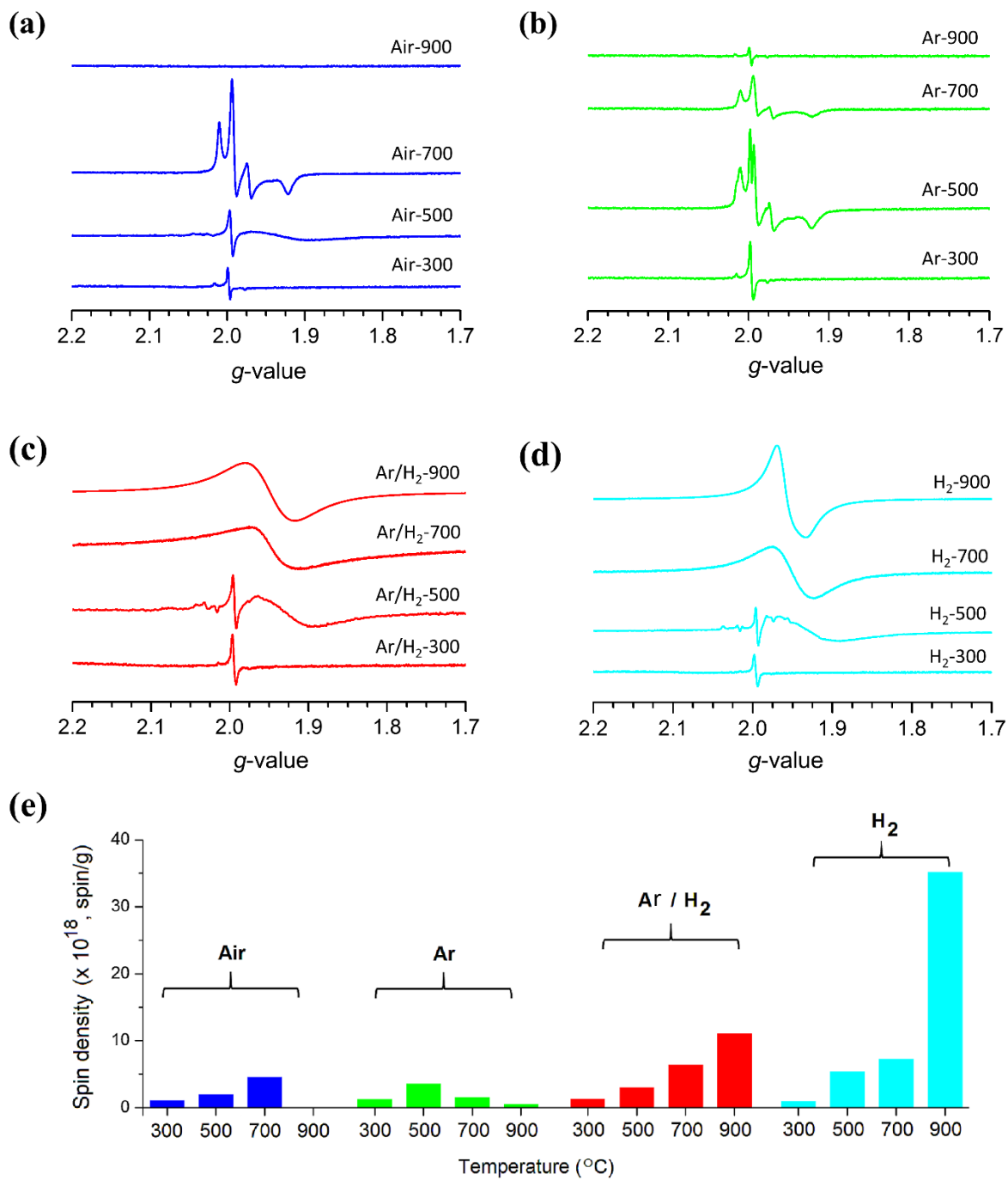

**Fig.3**